\documentclass[twocolumn]{aa}
\usepackage{graphicx}
\usepackage{txfonts}
\begin{document}
 
\title{First IBIS results on the high energy emission of Cyg X-2 
  \thanks
     {Based on observations with INTEGRAL, an ESA project with instruments
     and science data centre funded by ESA member states (especially the PI
     countries: Denmark, France, Germany, Italy, Switzerland, Spain), Czech
     Republic and Poland, and with the participation of Russia and the USA.}
 }

\titlerunning{First IBIS results on Cyg X-2}

\author{L. Natalucci\inst{1}, M. Del Santo\inst{1}, P.Ubertini\inst{1}, 
        F.Capitanio\inst{1}, M.Cocchi\inst{1}, 
	S. Piraino\inst{2}, A. Santangelo\inst{2}
          }

\authorrunning{Natalucci et al.}

\offprints{L. Natalucci}

\institute{CNR-Istituto di Astrofisica Spaziale e Fisica Cosmica, 
           Area Ricerca Roma 2/Tor Vergata, Via del Fosso del Cavaliere, 100
           00133 Roma, Italy\\
              \email{lorenzo@rm.iasf.cnr.it}
	 \and 
	   CNR-Istituto di Astrofisica Spaziale e Fisica Cosmica,
	             Sezione di Palermo, Via Ugo La Malfa 153,
		      90146 Palermo, Italy  \\
	     }

\date{Received; Accepted}

\abstract{
The bright low-mass X-ray binary Cyg X-2 was in the field of
view of the IBIS telescope during the early Cygnus region 
observations, executed during the INTEGRAL Performance
Verification Phase. 
The data presented are spanning about one week and cover 
the rising edge of one of the two peaks of the $\approx82$~day 
cycle of the Cyg~X-2 light curve. The IBIS data in the energy 
range 20-40~keV exhibit flux variation correlated with the 
{\em RXTE}/ASM light curve. Two different main exposures, 
separated by $\approx5$~days are found to be characterized by 
sensibly different spectra, with significant softening and 
higher X-ray luminosity in the second part, coincident with 
the long-term cycle peak.
At high energies, both measured spectra are very steep. The ratio 
of the 30-45~keV and 20-30~keV detected fluxes is $\approx0.30$, 
against a value of 0.95 expected for a Crab-like spectrum. No
positive detection exists for E~$\geq45$~keV, with a flux upper limit
($5\sigma$) of $\sim1.4\times10^{-10}$~erg~cm$^{-2}$~s$^{-1}$ 
in the 45-100~keV band.

         }
\maketitle

\section{Introduction}

Cyg~X-2 is a well known, and one of the best studied binary 
systems. Being a bright persistent source, many X-ray spectral and 
timing properties have been established and the main binary 
parameters determined during recent years observations. 
Cyg~X-2 is located at a distance 
of $\approx8$~kpc and is identified as a low-mass X-ray binary 
(LMXB), consisting of a weakly magnetized neutron star (NS) 
orbiting the late-type companion star V1341~Cyg with a 9.844
days orbital period (\cite{cowle79}). 
Evidence for the NS nature of the compact object 
comes from the detection of type~I X-ray bursts (\cite{smale98}) 
and from the optical measurement of the mass of the compact object, 
($1.78\pm0.23$)~${M}_{\odot}$ (\cite{orosz99}). The short time
duration ($\sim5$s) X-ray bursts detected from Cyg~X-2 are 
a challenge to the current theories. Short bursts are recognized as
typical of accretion rates lower than $\sim10^{-10}$~${M}_{\odot}$/yr
(\cite{kuulk02} and refs. therein)
so they are not expected from sources accreting at near-Eddington 
levels, like Cyg~X-2. Currently, short type~I bursts
have been observed in at least three high 
luminosity sources: Cyg~X-2, GX~17+2 (\cite{kuulk02}) and 
GX~3+1 (\cite{denha03}).
		   
\begin{figure}
\centering
\includegraphics[width=8.5cm]{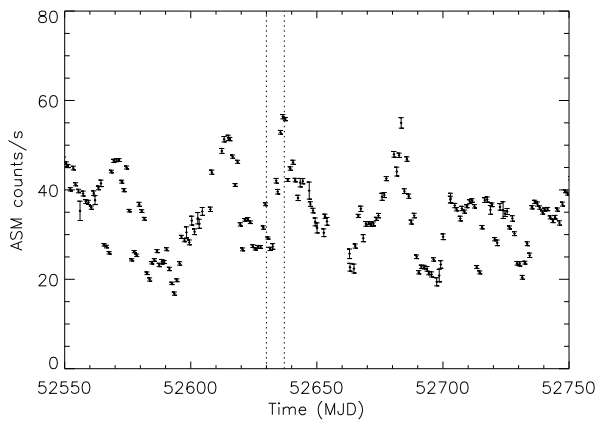}
\caption{A light curve of the 2-10 keV intensity of Cyg~X2 (data
         are courtesy of the {\em RXTE/ASM} team), showing the
	 double peak structure typical of the $\sim82$~days long
	 term cycle. Dotted lines indicate the epoch of
	 the IBIS observations reported here.
	           }
\label{Fig1}
\end{figure}

Cyg~X-2 has a bright, highly variable X-ray emission. Its X-ray 
light curve has a clear pattern, showing two peaks repeating 
periodically with a cycle of $\sim82$~days (\cite{vrtile03}, see 
also \cite{wijna96} for a previous determination), as is also
shown in Fig.~\ref{Fig1}. The X-ray 
luminosity changes regularly by a factor $\sim5$ to 10 in time 
scales of hours to days (see e.g.  \cite{disal02},
\cite{pirai02}). Associated to these luminosity changes are important 
spectral variations. When represented in terms of 
colour-colour diagrams (\cite{hasin89})
the spectral variations describe a Z-shaped track, with shape
and morphology depending on the source intensity
(\cite{kuulk96}). Distinct timing features (QPOs in the 
power spectrum) are found to be associated to the three different 
limbs of the Z (for a review see \cite{vklis00}). This behaviour has also 
been seen in other five sources common to the bright LMXB class:
GX~5-1, Sco~X-1, GX~17+2, GX~340+0 and GX~349+2 (Sco~X-2), 
forming the class of Z-sources. On short/medium time scales,
($\sim$~hours) the spectral softness if found to be  
positively correlated to the X-ray luminosity, consistently 
with the overall spectral behaviour of LMXBs in the X-rays.
However, the correlation of spectral/timing properties with 
accretion rate on long-term periods is still unclear. 
For example, the source may exhibit the same 
spectral softness and/or the same characteristic QPO 
frequencies while changing a factor of a few in X-ray 
luminosity. According to a recent interpretation 
(\cite{vklis01}), this is compatible with a long-time scale averaged 
response to variations of the accretion rate. On the other 
hand, as far as the spectral modelling is concerned, the 
most recent studies have converged to a 
generally agreed scheme, in which the soft component is 
described by a Planckian type law, i.e. blackbody or disk 
blackbody (see \cite{mitsu84}), whereas the hard component, 
luminous up to a few tens of keV is described by a thermal 
Comptonization spectrum (\cite{titar94}). This model, with the 
addition of an emission Fe~K$\alpha$ line and, in some cases, 
other lines for E~$\approx1$~keV (\cite{kuulk97}) is capable 
of providing a good description of the X-ray spectrum of Cyg~X-2 
(for a recent discussion, see \cite{cdone02} and refs. therein). 
In case of bright LMXBs, the Comptonization
spectrum is generally characterized by a relatively low
($\leq10$~keV) temperature of the electron plasma, whereas 
for atoll sources and most of the low luminosity 
LMXB X-ray bursters, ${kT}_{e}$~$\sim50$~keV
(\cite{barre00}). This is currently explained as an 
effect of cooling of the Comptonizing plasma, caused by a 
strong soft X-ray irradiation. 

Despite these latest developments, however,  
the picture of the origin of hard X-ray emission from Cyg~X-2,
as well for the other Z~sources, is not yet 
clear. Recent spectral measurements at high energies have 
detected a possible, further spectral component which is needed
to explain an observed hard tail. 
Modelled by a power law, this
hard tail has a photon spectral slope $\Gamma$ as hard as $\sim2$
(\cite{disal02}), or being in the range $\sim2.5-3.2$ 
(\cite{pirai02}). A similar hard tail has been also detected from 
other Z sources, namely GX~17+2
(\cite{disal00}), GX~349+2 (\cite{disal01}) and Sco~X-1 
(\cite{damic01},~\cite{stric00}). Hard tails in these sources are
found to be highly variable, with many cases of non-detection.
\begin{figure}
\centering
\includegraphics[width=8.5cm,height=6.0cm]{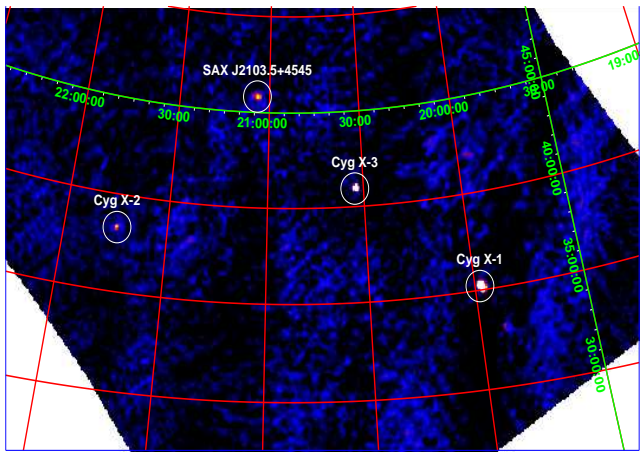}
\caption{IBIS/ISGRI mosaic image obtained from an
         observation of a field near to the Cygnus region, in the
	 energy range 20-30~keV. Cyg~X-2 is clearly
	 visible at the left, appearing at a significance level of 
	 23~$\sigma$. The other three sources visible,
	 Cyg~X-1, Cyg~X-3 and SAX~J2103.5+4545 are detected at
	 190, 126 and 24~$\sigma$, respectively. 
	           }
\label{Fig2}
\end{figure}

\section{Observations and data analysis}
\begin{table*}[]
\caption[]{Summary of the {\em INTEGRAL}/IBIS pointings used for the 
  analysis. For every pointing it shows the related 
  start time, duration, pointing
   coordinates RA,DEC (equinox 2000), and offset angle respect 
   to Cyg~X-2. 
  }
\tabcolsep=7mm
\hspace{1.1cm}
\begin{tabular}{c|c|c|c|c}

\hline 

\vspace{-3mm}\\
  Start time (MJD, days) & Duration (s) & R.A. (deg)  & Decl. (deg)   &  Offset (deg) \\
\hline

\vspace{-3mm}\\

  52629.63110 &  604 & 311.762 & 39.058 &   11.260  \\
  52629.63965 & 2164 & 311.663 & 39.116 &   11.335  \\
  52629.88000 & 2158 & 311.818 & 41.113 &   11.374  \\
  52629.90662 & 2157 & 311.988 & 43.109 &   11.748   \\
  52630.65271 & 2146 & 311.989 & 43.109 &   11.748  \\
  52630.67932 & 2283 & 311.818 & 41.113 &   11.374  \\
  52630.94812 & 2148 & 311.664 & 39.116 &   11.335   \\
  52631.21423 & 2147 & 311.818 & 41.113 &   11.374  \\
  52631.24084 & 2159 & 311.988 & 43.109 &   11.748  \\
  52636.66309 & 2246 & 325.154 & 49.879 &   11.580  \\
  52636.70850 & 2147 & 316.810 & 47.486 &   11.428  \\
  52637.22803 & 2095 & 311.901 & 42.113 &   11.521  \\
  52637.25476 & 2058 & 311.735 & 40.117 &   11.315  \\
  52637.49548 & 2100 & 311.593 & 38.120 &   11.444  \\
  52637.74109 & 2590 & 311.744 & 40.115 &   11.308  \\
  52637.77271 & 2092 & 311.907 & 42.115 &   11.517  \\
		 
\hline
\end{tabular}
\hspace{-2.2cm}
\label{tab:obs}
\end{table*}

The IBIS instrument on board {\em INTEGRAL} has been observing the 
Cygnus region extensively during the fall of 2002. In particular,
in observations when Cyg~X-1 was pointed off-axis, Cyg~X-2 was
in the partially coded field of view of the IBIS telescope. Within 
these observations we have selected IBIS/ISGRI data from 16
{\em science windows}, each corresponding to an exposure 
time of $\approx2150$~s. 

ISGRI is the soft $\gamma$-ray, CdTe detector of
{\em INTEGRAL}/IBIS, operative in the energy range 15-1000 keV with
an active area of 2600 cm$^{2}$. Its
on-axis sensitivity (5$\sigma$) in the range 15-50~keV is better than 
$\approx2$~mCrab with an exposure time of $10^{5}$~s. Due 
to the large field-of-view, ISGRI is an excellent monitor 
instrument for sources in regions extensively observed by 
{\em INTEGRAL}, in particular 
for sources at low Galactic latitudes. For a detailed description 
of IBIS and ISGRI we refer to \cite{huber} and 
\cite{lebru}, respectively. 

For the purpose of data selection, we required the
source offset $\le12$~deg from the telescope axis, in order to reduce 
the sensitivity losses and the unknown systematic effects which 
could arise, in this early phase, by very large offset angles. 
The selected observations are reported in detail in Table~1. 
The total IBIS exposure time for these pointings is 32.8~ks and is splitted 
in two main exposures, separated by a time gap of 5.5 days, whereas 
the offset of the source was always in the range 11.2-11.8 deg. The very
low spread of offset values is essentially due to the INTEGRAL 
observing strategy based on {\em dithering} around a given target
(\cite{winkl}), which was also adopted for part of the in-flight 
calibration observations.

\begin{figure}
\centering
\includegraphics[width=8.5cm]{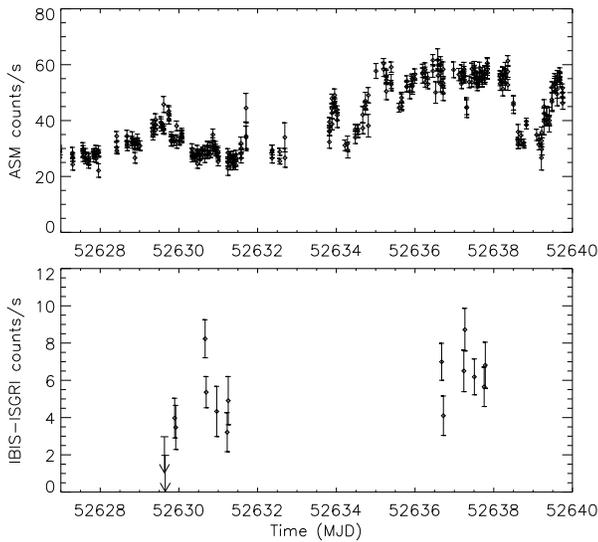}
\caption{Light curves of Cyg~X-2 obtained on November 2002, for
         {\em RXTE}/ASM (data are courtesy of {\em RXTE}/ASM team)
	 in the energy range 2-10 keV (top) and from IBIS/ISGRI
	 in the range 20-40 keV (bottom). The first two IBIS 
	 measurements are 2~$\sigma$ upper limits. Note that the 
	 unit used for ISGRI as counts/sec
	 are, to a certain extent, arbitrary (see text).
	           }
\label{Fig3}
\end{figure}

Cyg~X-2 was detected in all but the first two pointings. In Fig.~\ref{Fig2} 
we show a {\em mosaic} image in the energy band 20-30~keV, built 
using the total available exposure time. Cyg~X-2 is visible 
together with other bright, known sources: from right to left, 
Cyg~X-1, Cyg~X-3 and the transient SAX~J2103.5+4545, respectively 
(for the latter, we refer the reader to \cite{lutov03}). 
The data were processed 
by means of the IBIS off-line analysis S/W (\cite{gold1}), 
version 1.1. For each pointing we have extracted the count-rate flux measured 
after processing the related image. In Fig.~\ref{Fig3}, bottom panel, the light curve 
of the source in the energy band 20-40 keV is shown. The intensity values,
referred as ISGRI counts/s, are uncorrected for systematic effects due to 
the source offset and hence, cannot be compared directly to other 
on-axis measurements or to observations with significantly different
offsets. The reported intensity probably underestimates the 
ISGRI counts/s for an on-axis source by a factor $\sim1.5$. 
On the other hand, values of fluxes reported hereafter 
have been derived by comparing the ISGRI intensities
by the intensity measured from the Crab nebula during the calibration 
observations, at a close value of offset position ($9.6\deg$). So they
are corrected, to a large extent, from this type of systematics. The light 
curve obtained from {\em RXTE} All Sky Monitor (ASM) data in the energy 
band 2-10~keV is also shown. The IBIS intensities are found to be correlated 
to the ASM data, eventhough the high energy flux shows less variation. 
The period spanned by the data corresponds to the X-ray flux increasing 
approximately from the minimum to the maximum of one of the two peaks in 
the long-term cycle. Since the low energy data shows more intensity change, 
there is straightforward evidence from Fig.~\ref{Fig3}, for the 
spectrum to become softer in the second series of pointings.

For this reason and also due to the significant separation of the two
data sections, we consider two 
different exposures (see Table~2). In order to estimate hardness ratios 
we choose two contiguous energy bands, 20-30~keV and 30-45~keV. 
Reconstructed 
sky images were computed separately for the two exposures. 
From these images, and by 
cross-calibration with the Crab offset exposure we estimate 
average flux values, as reported in Table~2. In this preliminary phase 
of the calibration, we have allowed a 10\% systematic error in these 
flux estimates (for details see \cite{gold1}). 

The ratios of the 20-30~keV to the 30-45~keV count
rates are, after offset correction, ($2.89\pm0.56$) and ($3.33\pm0.59$) 
for the two observations. Therefore, the softening of the spectrum is evident 
only by comparison with the ASM data.

\begin{table*}[]
\caption[]{Flux values measured during the two main exposures (see
  text). For both exposures it shows: the related time interval, as 
  values of MJD-52000; the energy band, in keV; the measured flux,
  in units of $10^{-11}$~erg~cm$^{-2}$~s$^{-1}$. Errors in flux values
  allow a 10\% systematic error.
  }
\tabcolsep=5mm
\hspace{2.5cm}
\begin{tabular}{c|c|c}

\hline 

\vspace{-3mm}\\
 Time interval (days, MJD-52000) & Energy band (keV) & Flux ($10^{-11}$~erg~cm$^{-2}$~s$^{-1}$) \\
\hline

\vspace{-3mm}\\
 629.888-631.266  & 20-30 & $30\pm4$ \\
 629.888-631.266  & 30-45 & $9.2\pm1.9$  \\
 636.633-637.798  & 20-30 & $37\pm5$ \\
 636.633-637.798  & 30-45 & $9.7\pm1.9$ \\
\hline
\end{tabular}
\hspace{-4.5cm}
\label{tab:fluxes}
\end{table*}
\section{Conclusions}

We have reported on the first results of a monitoring of Cyg~X-2 performed
during $\sim10$ days in December 2002. Due to the large offset angle 
of the source and its steep spectrum, the observations were much limited in 
sensitivity. A joint analysis
of the IBIS/ISGRI and {\em RXTE}/ASM light curves reveals a softening
of the X-ray spectrum on a time scale of $\approx1$~week, corresponding
to the rising from one low intensity phase of the long term cycle of 
$\approx82$~days. During the softening, the 20-40~keV luminosities changed
only from $\approx3.5$ to $\approx4.2\times10^{36}$~erg~s$^{-1}$
(assuming a distance of 8 kpc) against a change of luminosity of a 
factor of $\sim2$ in the 2-10~keV band. During the second exposure, the
2-10~keV luminosity was $\sim10^{38}$~erg~s$^{-1}$ or even more. 
The hard X-ray spectrum is found to be very steep. The {\em luminosity} 
hardness ratios for the energy bands 30-45~keV and 20-30~keV are 
$\approx0.30$ or even less, indicating a spectral slope  
significantly greater than $\sim3.5$ (when this is approximated by a 
power law). Such a steep spectral slope is consistent with the shape of
the high energy tail of the thermal comptonization spectrum observed 
from Cyg~X-2. There is no evidence, instead, of a hard tail
extending beyond 40~keV, with a $5\sigma$ upper limit of $1.3\times10^{-3}$
ph~cm$^{-2}$~s$^{-1}$.

We note that the lack of detection of a hard tail is not contradictory
with other observations of such feature in the spectrum of Cyg~X-2, as 
this component could be variable or sporadic. Further monitoring and spectral
observations by INTEGRAL will help clarifying the nature of the high energy 
emission from this source.

\begin{acknowledgements}
The IBIS project is partially granted by Italian Space Agency (ASI). LN
is grateful to the IBIS engineering team at IASF/Rome, and in particular 
to M. Federici for efficient archive and software maintenance.  
\end{acknowledgements}

\end{document}